\documentclass[sigconf]{acmart}

\AtBeginDocument{%
  \providecommand\BibTeX{{%
    \normalfont B\kern-0.5em{\scshape i\kern-0.25em b}\kern-0.8em\TeX}}}

\copyrightyear{2022} 
\acmYear{2022} 
\setcopyright{rightsretained} 
\acmConference[FAccT '22]{2022 ACM Conference on Fairness, Accountability, and Transparency}{June 21--24, 2022}{Seoul, Republic of Korea}
\acmBooktitle{2022 ACM Conference on Fairness, Accountability, and Transparency (FAccT '22), June 21--24, 2022, Seoul, Republic of Korea}\acmDOI{10.1145/3531146.3533175}
\acmISBN{978-1-4503-9352-2/22/06}

\usepackage{algorithm2e}

\SetCommentSty{mycommfont}
\usepackage{listings}
\usepackage[colorinlistoftodos,prependcaption,textsize=tiny]{todonotes}
\usepackage{cleveref}
\usepackage{xspace}
\usepackage{blkarray}
\usepackage{booktabs}

\newcommand{\Smallset}{{Smallset}\xspace}
\newcommand{\Smallsets}{{Smallsets}\xspace}
\newcommand{\SsT}{{Smallset Timeline}\xspace}
\newcommand{\SsTs}{{Smallset Timelines}\xspace}
\newcommand{\Timeline}{Timeline\xspace}
\newcommand{\Timelines}{{Timelines}\xspace}
\newcommand{\sstool}{{\tt smallsets}\xspace}
\newcommand{\ssR}{{\tt smallsets R}\xspace}
\newcommand{\alttext}{{\textit{alt text}}\xspace}
\newcommand{\Alttext}{{\textit{Alt text}}\xspace}

\newcommand{\gr}{{U}\xspace}
\newcommand{\bl}{{E}\xspace}
\newcommand{\pu}{{A}\xspace}
\newcommand{\ye}{{D}\xspace}

\newcommand{\eat}[1]{}

\newcommand{\bX}{{\mathbf X}}
\newcommand{\bz}{{\mathbf z}}
\newcommand{\bA}{{\mathbf A}}
\newcommand{\bC}{{\mathbf C}}

\newcommand{\bhX}{{\mathbf {\hat X}}}

\begin{document}

\title{\SsTs: A Visual Representation of Data Preprocessing Decisions}





\author{Lydia R. Lucchesi}
\affiliation{%
  \institution{Australian National University \& CSIRO's Data61}
  \city{}
  \country{}
}
\email{Lydia.Lucchesi@anu.edu.au}

\author{Petra M. Kuhnert}
\affiliation{%
  \institution{CSIRO's Data61 \& Australian National University}
  \city{}
  \country{}
}
\email{Petra.Kuhnert@data61.csiro.au}

\author{Jenny L. Davis}
\affiliation{%
  \institution{Australian National University}
  \city{}
  \country{}
}
\email{jennifer.davis@anu.edu.au}

\author{Lexing Xie}
  \affiliation{%
  \institution{Australian National University \& CSIRO's Data61}
  \city{}
  \country{}
}
\email{lexing.xie@anu.edu.au}

\renewcommand{\shortauthors}{Lucchesi et al.}

\ccsdesc[500]{Human-centered computing~Visualization toolkits}

\keywords{data preprocessing, visualization, communication, open-source software, reflexivity}


\begin{abstract}

Data preprocessing is a crucial stage in the data analysis pipeline, with both technical and social aspects to consider.
Yet, the attention it receives is often lacking in research practice and dissemination.
We present the \SsT, a visualisation to help reflect on and communicate data preprocessing decisions.
A ``\Smallset'' is a small selection of rows from the original dataset containing instances of dataset alterations.
The \Timeline is comprised of \Smallset snapshots representing different points in the preprocessing stage and captions to describe the alterations visualised at each point.
Edits, additions, and deletions to the dataset are highlighted with colour.
We develop the R software package, \sstool, that can create \SsTs from R and Python data preprocessing scripts.
Constructing the figure asks practitioners to reflect on and revise decisions as necessary, while sharing it aims to make the process accessible to a diverse range of audiences.
We present two case studies to illustrate use of the \SsT for visualising preprocessing decisions.
Case studies include software defect data and income survey benchmark data, in which preprocessing affects levels of data loss and group fairness in prediction tasks, respectively.
We envision \SsTs as a go-to data provenance tool, enabling better documentation and communication of preprocessing tasks at large.

\end{abstract}

\maketitle

\section{Introduction}

Prior to an estimation task, data practitioners are often faced with difficult decisions about how to make their dataset functional \textit{for} the estimation task.
For example, one may need to decide how to deal with missing values to build a random forest classifier.
These data preprocessing decisions are important as they not only make the intended analysis possible but can also influence its outcome.
Their influence on estimation outcomes has been demonstrated, quantitatively, in the fields of fair machine learning \citep{Friedler2019}, natural language processing \citep{Denny2018}, and psychology \citep{Steegen2016}, to name a few.
Yet, in general it is less common to encounter meaningful detail about the preprocessing stage in discussions about research outputs, than it is to learn about how the data were collected and modelled \citep{Meng2021}.
Preprocessing decisions often remain tucked away in code---either inaccessible or difficult to parse, limiting our ability to interpret and replicate results.

Communicating and documenting data preprocessing is one aspect of data provenance, a broader concept referring to all aspects of dataset production. An influx of interest in data provenance in the machine learning community has led to work exploring how we might better record and utilise information about a dataset’s creation  \citep{DIgnazio2020, gebru2021datasheets, Hutchinson2021, Jo2020, Meng2021, Passi2017, Scheuerman2021}. Preprocessing is mentioned in the provenance literature, but because there are many aspects of provenance, it receives limited attention. Meanwhile, the field of information visualisation has produced tools to study data provenance and its effects. Some support visualisation of the entire data pipeline \citep{Wang2020}, data lineage \citep{Cui2000}, or data flow \citep{Yang2020}. Others are interactive \citep{Bors2019, Callahan2006, Niederer2017} or animated \citep{Pu2021, Khan2017}. To the best of our knowledge, none of the existing tools focus on visualising the \textit{decisions} made during preprocessing in a way that is static and compact. We choose to focus on this.

We present the \SsT (or \Timeline), a visualisation of data practitioners’ preprocessing decisions (\Cref{sec:design}). 
A \Smallset is a small collection of rows from the dataset containing examples of data alterations.
Rows are selected by random sampling or one of the proposed optimisation algorithms (\Cref{sec:smallset_selection}).
The \Timeline is comprised of \Smallset snapshots representing different points in the preprocessing steps and captions to describe the alterations visualised at each point. 
Edits, additions, and deletions to the data are highlighted with colour.
It is a static, compact visualisation designed to be useful for both \Timeline creators and readers (\Cref{table:functions}). A \Timeline creator is one who makes a \SsT to reflect on and communicate their decisions. A \Timeline reader is one who views it to understand, evaluate, and/or replicate the preprocessing steps. We present the \sstool R package (\Cref{sec:software}), which is used to produce all \Timelines in this work, including those in the case studies (\Cref{sec:case_studies}). The first case study uses \SsTs to document decisions related to varying amounts of data loss in software defect data from the NASA Metrics Data Program. The second case study explores American Community Survey benchmark datasets from the \texttt{folktables} tool \citep{Ding2021} and the subtle downstream effects of combining different filtering and threshold decisions.

 The main contributions of this work are: 
\begin{itemize} 
  \setlength\topsep{-.1em}
    \item The \SsT, a static, compact visualisation to communicate data preprocessing decisions.
    \item 
    The open-source package \sstool \footnote{\href{https://github.com/lydialucchesi/smallsets}{\color{blue}https://github.com/lydialucchesi/smallsets}}, for producing \SsTs for R and Python preprocessing scripts.
    \item Two case studies, in which \SsTs document preprocessing decisions that affect comparability of results from different studies as well as dataset imbalance and group fairness in machine learning tasks.
\end{itemize}

\section{Related Work}
\label{sec:rel_work}

We review several areas of related research that motivated and inspired the creation of the \SsT. 
These areas include 1) studying the effects of preprocessing decisions on outcomes from data analytics tasks, 2) documenting data provenance, and 3) visualising data provenance information.

There is rarely a clear-cut preprocessing route for practitioners to follow. 
Instead, practitioners must make decisions about how to prepare data for analyses. 
Research about \textbf{preprocessing effects} investigates if study outcomes are sensitive to these decisions. 
For example, \citet{Friedler2019} uncover dependence between performance of fairness-enhancing algorithms and preprocessing choices. 
\citet{Blocker2013} introduce the concept of multiphase inference for preprocessing to obtain better estimators. 
\citet{Steegen2016} propose multiverse analyses, in which a dataset is prepared a number of reasonable ways for estimation. 
They demonstrate with a psychology case study that estimation outcomes can be sensitive to differences in data preprocessing. 
Similarly, \citet{Denny2018} show that, in the preparation of political texts for unsupervised learning tasks, ``under relatively small perturbations of preprocessing decisions---none of which were \textit{a priori} unreasonable---very different substantive interpretations would emerge'' [p.~187]. 
Conversely, in an experiment by \citet{Shirk2017}, three participants remove artefacts from electroencephalogram (EEG) data, and despite varied approaches, the results remain stable. 
However, generally speaking, these works do provide compelling quantitative evidence that preprocessing \textit{can} shape the trajectory of an analysis. 
From this literature, we can conclude that careful review and strong communication of preprocessing decisions are important.

The \textbf{documentation of data provenance} is a growing area of interest for the machine learning and natural language processing communities.
The goal is to record important information about a dataset and support informed use of data and models. Datasheets for datasets \citep{gebru2021datasheets}
, model cards \citep{Mitchell2019}, FactSheets \citep{Arnold2019}, data statements \citep{Bender2018}, and the Dataset Nutrition Label \citep{Holland2018} are proposed templates and frameworks for recording information about a dataset, including changes made to it.
For example, Question 33 in the datasheets for datasets template asks: ``Was any preprocessing/cleaning/labeling of the data done (for example, discretization or bucketing, tokenization, part-of-speech tagging, SIFT feature extraction, removal of instances, processing of missing values)?'' \citep[p.~90]{gebru2021datasheets}.
These documentation techniques aim to be comprehensive in their coverage of provenance information.
Recorded preprocessing details are just one part of the documentation. 
With the \SsT, we hope to contribute to this research area with a technique focused exclusively on the topic of data preprocessing.
Furthermore, we explore the pairing of text \textit{and} visuals to describe data alterations.

Next, we discuss \textbf{visualisations for data provenance} that include information about preprocessing or enable end users to study data transformations. 
\citet{Wang2020} use data comics \citep{Zhao2015} to describe an analytical process, with some panels dedicated to data transformations. 
\citet{Cui2000} propose a data lineage tracing algorithm and exploration tool. 
DQProv Explorer \citep{Bors2019} and VisTrails \citep{Callahan2006} are multi-view interactive visualisation systems providing insight into the transformations undergone by a dataset. 
TACO \cite{Niederer2017} is another interactive system offering several visual summaries for data table comparisons across time. \citet{Khan2017} develop ``data tweening,'' which involves animating the transformations occurring between two database queries. 
A ``datamation'' \cite{Pu2021} animates plotted data points to showcase restructuring tasks, while \citet{Yang2020} propose \textit{fair}-DAGs for identifying bias in preprocessing pipelines. 
These tools convey provenance information using sketches, interactivity, animation, and directed acyclic graphs (DAGs). 
We focus solely on preprocessing and propose a static timeline of steps. 
The design is intended to be simple and practical. 
We discuss the \SsT in detail in \Cref{sec:design} after outlining the role of this visualisation in \Cref{sec:roles}.


\section{The Role of Smallset Timelines}
\label{sec:roles}

\begin{figure*}[!t]
\centering
\includegraphics[width=6in, keepaspectratio]{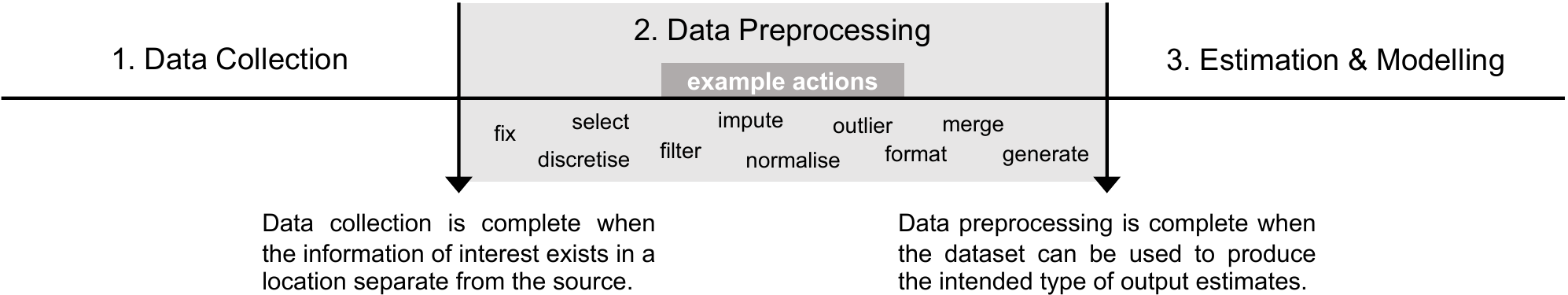}
\caption{Operational definition for data preprocessing. Boundary definitions distinguish preprocessing from neighbouring stages in the data pipeline. In practice, some iteration between stages may be necessary.}
\label{fig:definition}
\end{figure*}

\begin{table*}[!t]
\small
\caption{Design goals for Smallset Timelines - including users, utilities, and the corresponding format variations.}
\label{table:functions}
\begin{tabular}{lp{4cm}p{4cm}p{4cm}}
    \toprule & \textbf{Reflect} & \textbf{Replicate} & \textbf{Comprehend \& Evaluate} \\ \midrule
    \vspace{.1cm}
    \textbf{User: action} & Data practitioner: create & Data practitioner: read & Target audience: read \\
    \vspace{.2cm}    
    \textbf{Outcome} & 
    Asks practitioners to recount their decision-making process in the Timeline captions, encouraging reflection for the data preprocessing stage. & 
    Provides visual examples and written descriptions of preprocessing steps performed in a programming language. Documents information in a stable format that can be saved. & 
    Provides an accessible preprocessing narrative with enough information for decisions to be understood and assessed. Highlights the role humans play in data production. \\
    \vspace{.2cm} 
    \textbf{Presentation} & 
    Reflection occurs while writing captions for the \Timeline. & 
    \Smallset has more rows. \newline \Timeline has more snapshots. \newline Captions are detailed. &
    \Smallset has fewer rows. \newline 
    \Timeline has fewer snapshots. \newline 
    Captions are succinct. \\
    \vspace{.2cm} 
    \textbf{Example} & 
    \Cref{fig:gray_general,fig:gray_replication,fig:census} & 
    \Cref{fig:gray_replication} & 
    \Cref{fig:gray_general,fig:census} \\ \bottomrule
\end{tabular}
\end{table*}

In this section, we first clarify the meaning of \textit{data preprocessing} used in this work.
We then define the roles of the \SsT for different users and goals.
\textit{Data preprocessing} is a commonly used term in the research and practice of data science, but the term carries a diverse set of meanings that vary with context and audience. Following are three example views of preprocessing, ranging from specific to general.
For text data, \citet{Denny2018} view preprocessing as the set of ``decisions about how words are to be converted into numbers'' [p.~168].
Focused on the role of preprocessing in data mining, \citet{Garcia2015} define it in terms of two broad task categories, including \textit{data preparation} and \textit{data reduction}, and the sub-tasks they encompass, e.g., \textit{data cleaning} or \textit{feature selection}.
One notable challenge of this approach is developing a classification scheme that is comprehensive.
In a general overview of data preprocessing, \citet{Famili1997} simply define it as `` all the actions taken before the actual data analysis process starts'' [p.~5].
We build on this last conceptualisation and adhere to a minimalist conception of data preprocessing, focusing on its boundaries with other stages in the data pipeline.

Figure~\ref{fig:definition} pictures a three-stage pipeline, consisting of 1) data collection, 2) data preprocessing, and 3) estimation and modelling. We consider data collection complete when the information of interest exists in a location separate from the source and data preprocessing complete when the dataset can be used to produce the intended type of output estimates.
These boundaries delimit the beginning and end of preprocessing.
Actions altering the dataset within these boundaries are considered preprocessing. \Cref{fig:definition} includes some example actions, and more examples can be found in \citet{Famili1997}, \citet{Kasica2020}, and \citet{Luengo2020}, to name a few.
This conceptualisation of preprocessing is robust to variability in the location of data operations across analyses. For instance, inference can generate a new feature or produce the final result.
Here, if the inference task changes the dataset to facilitate the analysis, it is preprocessing.
Although our definition implies a linear and pre-defined analytic strategy, we acknowledge and account for exploratory or iterative processes through the \textbf{resume marker} feature introduced in \Cref{ssec:op_features}.
However, we omit a full discussion of this feature due to space constraints. 

We develop the \SsT to capture the nature of preprocessing actions altering the dataset. 
We create the visualisation and tool to serve several functions, outlined in \Cref{table:functions}. 
One function is to support reflection on preprocessing decisions by the person who made the decisions, or the \Timeline creator. 
There is a growing call to incorporate reflection---especially reflexivity \citep{Elish2018, Tanweer2021, Miceli2021}---into data science work to acknowledge the context and subjectivities involved in it. 
Asking practitioners to recount their decision-making process in the \Timeline captions aims to promote reflection about the preprocessing stage. 
The second function is to support replication of the steps by other researchers. 
Reproducibility is considered a cornerstone of science \citep{Nosek2021}, and being able to replicate preprocessing is an essential component of reproducing data-based results. 
The third function is to support comprehension among a \Timeline creator's target audience. 
Given the importance of preprocessing decisions, as established in \Cref{sec:rel_work}, getting preprocessing decisions out of code and into an accessible and practical format is crucial for making these decisions legible and thus open to evaluation. 
Next, we describe the \SsT design and how it affords these socio-technical functions.

\section{Smallset Timeline Design}
\label{sec:design}

The design of any tool enacts priorities through its ``affordances,'' or how technical features interplay with human users to produce socially meaningful effects \citep{Davis2020}. 
In this section, we describe the design of the main visual artefact, the \SsT, noting how specific design choices relate to intended use-functions, including reflection, replication, and comprehension/evaluation (\Cref{table:functions}).
 
A \SsT has three basic components: a \textit{\Smallset} consisting of a small subset of data to illustrate preprocessing decisions, \textit{snapshots} that each visualise one or more preprocessing decisions, and \textit{captions} that describe changes made to the data (\Cref{ssec:components}). 
The timeline also has four enrichment design features: \textit{printed data}, \textit{missing data tints}, \textit{ghost data}, and \textit{resume markers} (\Cref{ssec:op_features}).
We also generate alternative narratives (\textit{alt text}) for \SsTs for those with visual impairments (\Cref{ssec:alt_text}). 
Throughout this section, we use a synthetic dataset and preprocessing scenario to illustrate various design components and their functions. 
The synthetic dataset consists of 100 rows and 8 features. 
The main preprocessing steps are 1) filtering rows, 2) dealing with missing data, and 3) generating a new feature. 
More information about the synthetic dataset and preprocessing scenario can be found in \Cref{app:synthetic}.

\subsection{Key components}
\label{ssec:components}

A \textbf{``\Smallset''} is a small collection of observations featuring examples of data alterations occurring in the dataset of interest. 
In this section, we assume these observations are given in order to focus on the visual elements of the design component. 
\Smallset selection criteria and algorithms are discussed in \Cref{sec:smallset_selection}. 
The design goal is to create a small object that can demonstrate preprocessing steps at a manageable scale for comprehension and figure production. 
A \Smallset contains approximately 5-15 observations to keep the visualisation compact. 
The current version of the \SsT tool works for tabular data only, meaning the \Smallset also has this table format. Each observation is a row. 
Each attribute is a column, and there is no nested data structure (e.g., lists or other key-value structures) in a cell.

\begin{figure*}[!t]
\centering
\includegraphics[width=\linewidth, keepaspectratio]{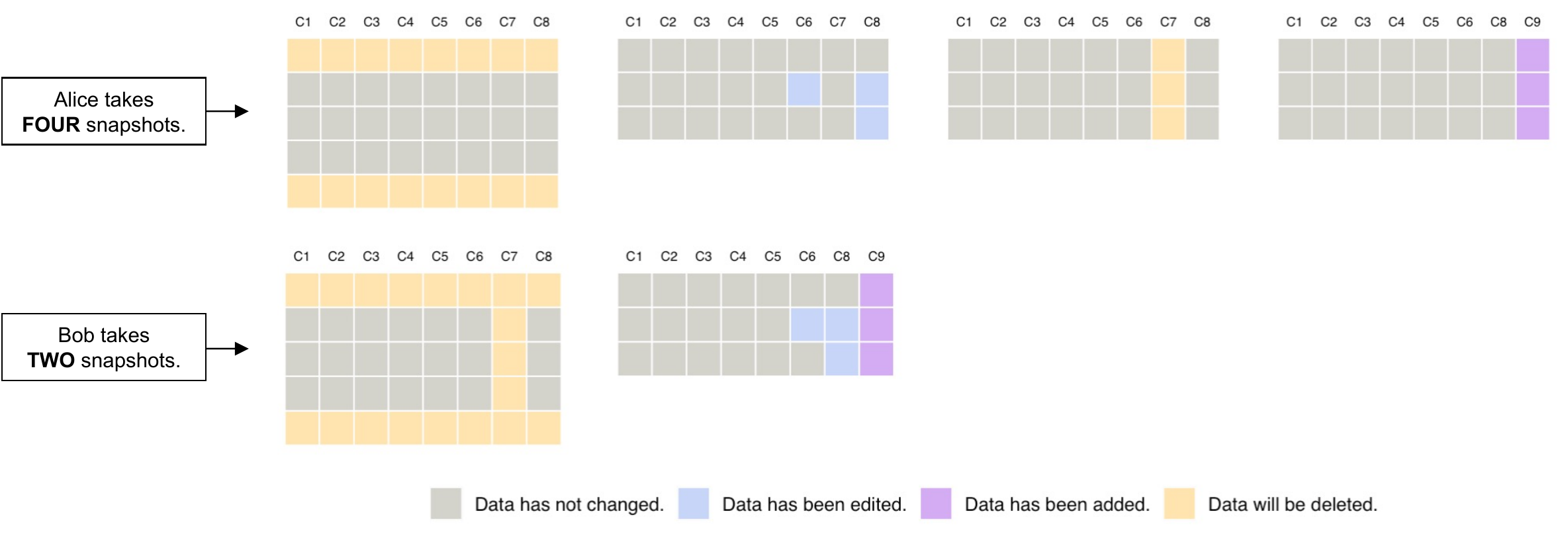}
\caption{Diagram showing discretion in snapshot point selection. Alice generates snapshots for each step. Bob combines preprocessing steps in snapshots. Snapshots are based on synthetic data (\Cref{app:synthetic}).}
\label{fig:snapshots}
\end{figure*}

Small empty tables have long been used in the programming community to explain coding commands for data manipulation. 
For example, the cheat sheet for the R \texttt{dplyr} package \cite{RStudio2021} uses little (empty) tables and colour to visually explain to data scientists what happens to the data object when a \texttt{dplyr} command is applied to it. 
With the \Smallset, we employ the same technique. 
It allows \Timeline creators to demonstrate to \Timeline readers what happens to a dataset as a result of their preprocessing decisions. 
A \Smallset is not limited to showing one operation at a time but can instead show multiple programming steps at once (e.g., \Cref{fig:snapshots}). 
Providing real examples of dataset changes, in a convenient viewing format, is one way to make preprocessing transparent to Timeline readers.

\textbf{Snapshots} are pictures of the \Smallset table at a particular moment in the data preprocessing steps. 
Snapshots break the process into digestible pieces and are plotted sequentially in a timeline to mirror the sequence of programming instructions used to implement a data preprocessing strategy. 
The first snapshot shows the data prior to any preprocessing, while the last presents it fully preprocessed. 
Snapshots in-between represent intermediary points, selected by the \Timeline creator (by simply inserting structured comments, see \Cref{sec:software}). 
Snapshots can be arranged in a single row or across multiple rows.

The system uses a set of colours to highlight data changes in a snapshot. 
The colours represent general changes undergone by a dataset: 1) it gets bigger, 2) it gets smaller, or 3) it stays the same size, but the contents change. 
In short, it is a colour scheme distinguishing between data additions, deletions, edits, and unchanged data. 
We limit the number of colours to four to minimise consultation with the colour legend while reading a \Timeline. \Timeline creators can choose a four-colour palette consistent with the visual style of their document, and colourblind-friendly palettes are available in the \sstool package. 
The colour for a data change not appearing in the \Timeline is dropped from the legend (e.g., \Cref{fig:gray_general}). 
We leave experimenting with the number and type of labelled changes, as well as the option to assign colours to specific operations, as future work. 

\Timeline creators are expected to exercise discretion in snapshot-taking based on their goals and presentation format. 
In \Cref{fig:snapshots}, \Timeline creator Alice chooses to take snapshots showing exactly one operation at a time. 
As noted in \Cref{table:functions}, this type of approach emphasises the effects of each operation and helps prepare documentation for replicating the data preprocessing tasks. 
Alternatively, another \Timeline creator, Bob, groups related operations together as a composite preprocessing step. 
This type of approach aims to convey the conceptual outline rather than the details of preprocessing. 
It is suited to mediums in which space and reader attention span are limited, such as a research article, white paper, or blog post. 
It should be noted that if a data point has been altered more than once since the last snapshot, the cell colour will reflect the most recent change, i.e., one operation becomes hidden behind another. 
However, we choose to prioritise simplicity and minimise visual clutter.

\textbf{Captions} accompany snapshots to provide information about the alterations visualised in the Smallset. 
\Timeline creators are responsible for providing the captions (by populating a caption template, see \Cref{sec:software}), which should  supply  \Timeline readers with information that enhances their understanding of the process. 
This text is generally located beneath snapshots but could be placed to the side, if a Timeline is arranged vertically.

At the most basic level, a caption says what was done in the preprocessing step. 
The colour categories for data changes are broad, so a caption allows the exact nature of the change to be stated. 
From there, the caption can be upgraded to also explain why it was done. \Timeline creators can use the caption space to defend and discuss their preprocessing decisions. 
Explaining why is especially important if a decision deviates from a preprocessing norm in one’s field. 
In some instances, it may be necessary to also specify how it was done. 
This part can be essential for \Timeline readers trying to replicate the preprocessing steps.

The caption style will depend on the purpose of the \Timeline. 
To caption appropriately for general comprehension (\Cref{table:functions} column 3), jargon is avoided, and the text is pared back to the most relevant parts to prevent information overload. 
Caption 1 in \Cref{fig:gray_general} provides an example of a simple caption for general comprehension: \textit{Remove columns that have the same value for every row because they do not provide any information for modelling.} 
For the purpose of replicating data preprocessing tasks (\Cref{table:functions} column 2), captions may be detailed, include jargon, and reference preprocessing code. 
Those reproducing the steps likely have some familiarity with the topic, such that the amount and type of information are not overwhelming. 
The captions in \Cref{fig:gray_replication} are an example of captioning to enable replication. 
For example, the step 4 caption lists the integrity rules used to check for implausible values.

\subsection{Enrichment features}
\label{ssec:op_features}

\begin{figure*}[!t]
\centering
\includegraphics[width=5.5in, keepaspectratio]{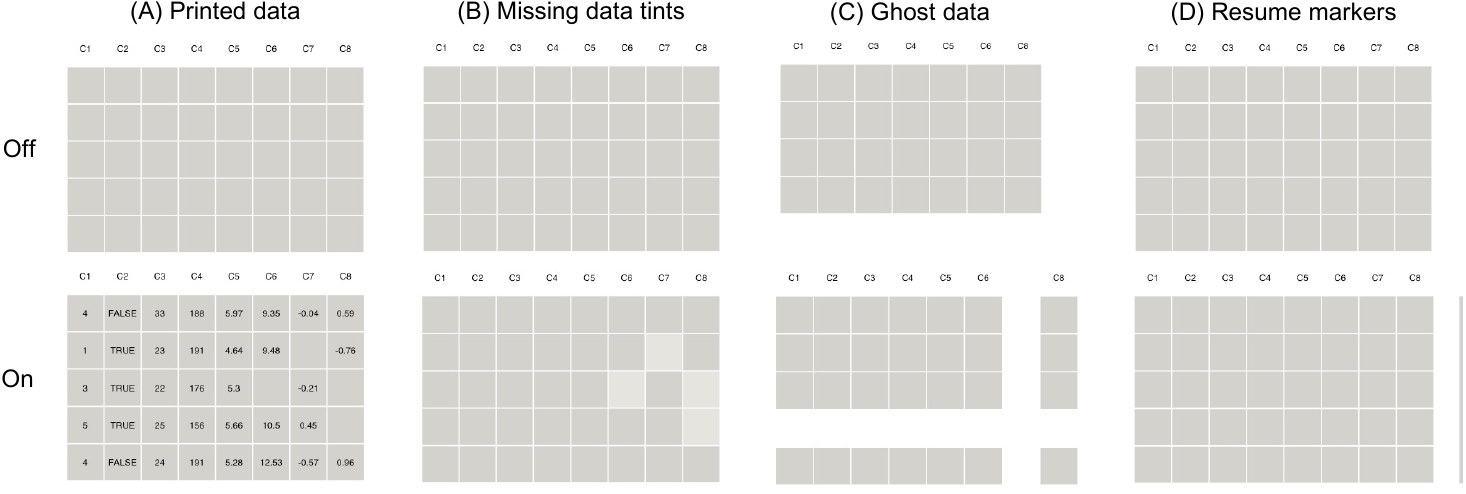}
\caption{Overview of Smallset enrichment features. See \Cref{ssec:op_features} for descriptions. Snapshots are based on synthetic data (\Cref{app:synthetic}).}
\label{fig:op_features}
\end{figure*}

The \SsT is designed with four enrichment features. Their use is at the discretion of \Timeline creators and should depend on data privacy as well as audience and goal (\Cref{table:functions}). A visual overview of the features is in \Cref{fig:op_features}.

\textbf{Printed data} (\Cref{fig:op_features} column A) can be included in Smallset tables for a glimpse of the data and real examples of how it changes between steps. 
If reading a \SsT to help reproduce a dataset, printed values provide a chance to compare values between the original dataset and the reproduced version. 
Even if preprocessing code appears to run successfully, having the printed values might verify that it still does what the author intended. 
If data are not publicly available, the \Timeline can be configured without printed data. 
Note that omitting the data does not necessarily guarantee data privacy. 
We hope to address Smallset data privacy in future work for applications with sensitive data.

The ``missing value shadow'' \citep{Swayne1998}, or ``shadow matrix'' \citep{Tierney2021}, is a visual technique that contrasts light and dark to reveal missing values in a table of data. 
Utilising the concept, the \SsT tool offers the option to indicate missing values in a Smallset with \textbf{missing data tints} (\Cref{fig:op_features} column B). 
A subtle tint makes the issue noticeable without diverting attention from other visual elements. 
It also has metaphorical value, as if part of the colour is missing. 
If values are imputed, the table cells are not filled with tints in the following snapshots.

When deletion is demonstrated in the \Smallset, the \Smallset table naturally shrinks. When this happens in a \Timeline, it can become difficult to track data points across the \Timeline as they shift in space relative to each other. 
The \textbf{ghost data} enrichment feature (\Cref{fig:op_features} column C) provides the option to plot blank (``ghost'') rows and columns where data have been deleted. 
The position of each table cell is then maintained throughout the \Timeline, and data can be readily traced across it (e.g., \Cref{fig:gray_general}, \Cref{fig:gray_replication}, and \Cref{fig:census}).

\textbf{Resume markers} (\Cref{fig:op_features} column D) are the fourth enrichment feature. 
A resume marker is a vertical line placed between two snapshots to denote that preprocessing stopped to begin the estimation task and then subsequently restarted for additional data alterations to be made. 
This feature is designed to be of use when discussing iterations in more exploratory analyses or unexpected issues that necessitate modifying the initial estimation plan. 
In the latter case, this enrichment feature is not meant to condone any type of data dredging but allow \Timeline creators to be transparent about unforeseen roadblocks, resulting in multiple attempts at estimation. 
An example \SsT with a resume marker can be found in \Cref{app:synthetic} (\Cref{fig:resume}).

\subsection{Alt text}
\label{ssec:alt_text}
In the present work, we argue that visualisation is a good way to make preprocessing information accessible. 
However, visualisations are not accessible to people with visual impairments unless there is \textit{alt text}. 
Therefore, we develop an \textit{alt text} template (available in \Cref{app:alt_template}) for generating text descriptions of \SsTs. 
The software tool, discussed in \Cref{sec:software}, automatically populates the template and saves the output in a text file. 
When populated, it details the \Timeline title, snapshot count, colour legend, and individual snapshots. 
This output can be manually modified for clarity and readability and included alongside figures. 
An example of the automated \textit{alt text} as well as the manual edits a practitioner could make prior to dissemination are included in \Cref{app:alt_output}.

\section{Smallset Selection}
\label{sec:smallset_selection}

Section~\ref{sec:design} assumes a small set of rows from a tabular dataset is given. This section discusses three strategies to automatically select these rows from the original dataset.
There are two main criteria for selecting a Smallset.
{\em Preprocessing coverage} tries to ensure that at least one row in the Smallset is affected by each preprocessing step, so that all snapshots in the timeline have a visible change. 
{\em Visual variety} aims to select a set of rows that are affected by the set of preprocessing steps differently, so as to represent a range of changes from preprocessing.
Although manual selection is an option, it is subject to cherry-picking, which may result in a misleading visualisation. The tool offers three methods for automated selection to bolster integrity of the visualisation presented.

Random sampling, the first automated selection method, may achieve \textit{preprocessing coverage} and \textit{visual variety} if preprocessing operations are widespread throughout a dataset. However, when the number of \Smallset rows is low, and when some preprocessing operations affect only a small number of rows in the original data, neither \textit{preprocessing coverage} nor \textit{visual variety} are guaranteed, which makes other automated selection methods desirable.

\subsection{Two optimisation problems}

Automatic selection algorithms require two additional data representations generated from the preprocessing steps: the {\em coverage indicator matrix}, $\bC$, and the {\em visual appearance matrix}, $\bA$. Denote the original dataset $\bX$ as an $N{\times}M$ matrix, with $x_{ij}$ being the data value in the $i$-th row and $j$-th column. Data matrix $\bX$ goes through $h=1,\ldots,H$ preprocessing steps, $f_1, \ldots, f_H$, resulting in a processed data matrix after each step $\bhX_h = f_h \ldots f_1(\bX)$.
The binary coverage matrix $\bC \in \{0, 1\}^{N \times H}$ is sized by the $N$ data points and $H$ preprocessing steps. Each element $c_{ih}$ is 1 {\it iff} the $i^{th}$ row is altered by preprocessing step $f_h$, 0 otherwise.
The appearance matrix $\bA \in \mathcal{R}^{N' \times M'}$ is the size of the original data matrix plus any rows\slash columns added.
Its elements $a_{ij} \in \{'U', 'E', 'A', 'D'\}$ (corresponding to {\em unchanged, edited, added, deleted}, respectively) encode the last change that a data cell undergoes from the original data matrix $\bX$ to the final data matrix $\bhX_H$.
Example {\em coverage indicator} and {\em visual appearance} matrices for the synthetic dataset are available in \Cref{app:score_matrices}.

We use these data structures to set-up two optimisation problems (Problem 1 and Problem 2 shown in \Cref{table:opt_selection}) for selecting a \Smallset of size $K$.
Problem 1 accounts for {\em preprocessing coverage} only. The output of this selection problem is an indicator vector $\bz \in \{0, 1\}^N$, with $z_i$ being 1 if row $i$ is selected, 0 otherwise. The first constraint ensures that exactly $K$ rows are selected out of the original $N$ rows. In the second constraint, the left hand side computes the number of rows that preprocessing step $h$ affects, and we require this to be greater than 0 for each step. 
This is an integer linear program solved using the Gurobi~\cite{gurobi} optimisation software.
In other words, the {\em coverage} problem tries to satisfy the two constraints without any additional objective (hence the $\max 1$ term in \Cref{table:opt_selection}). We tried maximising the number of changes shown, but that led to solutions that favour rows with many changes, that may all be similar to each other -- which motivates the {\em visual variety} criterion and Problem 2.
 
Problem 2 additionally accounts for \textit{visual variety}. This requires a pre-calculated $N{\times}N$ distance matrix $Q$ containing the hamming distance between the appearance vector of any two rows. That is, $q_{il} = \sum_j d(a_{ij},a_{lj})$, with distance function $d(\cdot,\cdot)$ being 0 if the two values are the same, 1 otherwise.
The objective function $\bz^\top Q \bz$, therefore, computes the total pair-wise hamming distance among the selected rows. The two constraints remain the same as Problem 1. This is an integer quadratic problem, also solved with Gurobi~\cite{gurobi}.

\begin{table}[h]
\footnotesize
\caption{Two optimisation problems for Smallset selection.}
\label{table:opt_selection}
\begin{tabular}{p{3.9cm}  p{3.9cm} }
      {\small Problem 1 - Coverage} & {\small Problem 2 - Coverage + Variety} \\
      {$\!\begin{aligned}
      \max_\bz \quad & 1 \\ 
      \textrm{s.t.} \quad & \sum_{i=1}^N z_i = K \\
      & \sum_{i=1}^N z_i c_{ih} > 0, ~\forall h=1,\ldots, H
      \end{aligned}$}
      & 
      {$\!\begin{aligned} %
               \max_\bz \quad & \bz^\top Q \bz \\    
               \textrm{s.t.} \quad & \sum_{i=1}^N z_i = K \\
      & \sum_{i=1}^N z_i c_{ih} > 0, ~\forall h=1,\ldots, H
               \end{aligned}$}
\end{tabular}
\end{table}

\begin{figure*}[!t]
  \centering
  \includegraphics[width=\linewidth]{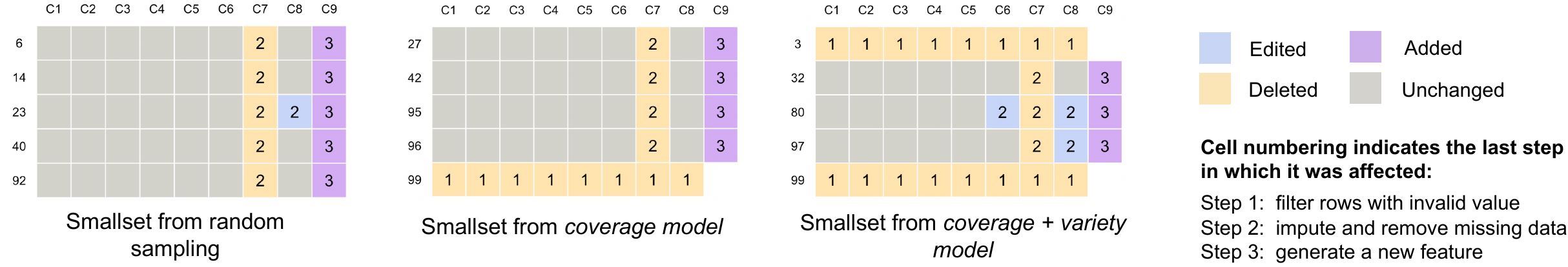}
  \caption{Smallsets selected by random sampling (left), coverage model (middle), and coverage+variety model (right) on the synthetic dataset (\Cref{app:synthetic}). Shown here is one snapshot with accumulated changes (indicated by cell color) across the three processing steps (indicated by the numbering of the cell). Row numbers refer to those in the original dataset.}
  \label{fig:selection_result}
\end{figure*}

Figure~\ref{fig:selection_result} illustrates the three different approaches for selecting $K=5$ rows: random selection, selection that prioritises coverage, and selection that prioritises coverage and variety.
We can see that random sampling misses a row affected by Step 1.
While the solution from Problem 1 satisfies the constraints of covering Steps 1, 2, and 3, the first four selected rows underwent the same preprocessing steps.
The solution from Problem 2, thanks to the {\em visual variety} criterion, selects three rows affected by Step 2 and 3 differently.
Another desirable by-product of {\em visual variety} is having one row with minimal changes included (row 32).

\subsection{Discussion}
\subsubsection{Novelty} The two proposed optimisation algorithms are similar in spirit to known combinatorial problems on sets~\cite{cormen2009introduction}, but the particular formulation incorporating preprocessing workflow and visual appearance criteria is new. 
Our Smallset selection algorithms are distinct from known subset selection approaches, as our objective functions are not submodular~\cite{wei2015submodularity}. 
Our solution relies on auxiliary data generated from the preprocessing steps and does not need to cluster the input~\cite{daszykowski2002representative}, noting clustering would require preprocessing having been completed.

\subsubsection{Comment on running time} Despite being combinatorial optimisation problems, we obtained solutions for Problems 1 and 2 for the synthetic dataset in a few seconds. 
Problem 2 generates visually more desirable outputs, at the cost of needing to precompute and optimise with a distance matrix $Q$ that is quadratic in the number of rows.

\subsubsection{Potential problem variants} 
One may wonder whether this methodology could be used to select a subset of columns, e.g., 41 columns in the MDP CM1 dataset (\Cref{ssec:mdp_understanding}) is clearly too many for the visualisation.
Indeed, one can envision formulating variants of the objective and constraints from the coverage and appearance matrices for each column. 
We leave this as future work.
One may also wonder whether other objectives are needed, such as {\em representativeness} -- for more selected rows to reflect changes that often occur, and vice versa.
This is possible, e.g., by minimising the difference between the fraction of changes in sampled rows and those for the whole dataset in Problem 1.
However, we note that computing reprensentativeness on a small subset is prone to statistical noise and exclude it from our primary criteria.

\subsubsection{Do \SsTs have to be small?}
We recommend that \Smallsets be 5-15 rows and \Timelines be 2-10 snapshots, due to the cognitive limits of the \Timeline reader and the constraints of having readable visualisations within limited page or screen space.
Multiple snapshots with a larger \Smallset can be used to accommodate longer chains of operations that affect a large dataset in diverse ways. 
In these scenarios, the choice is left to \Timeline creators to trade-off between a large visualisation with many details or a small one with fewer details.

\section{The smallsets R Package}
\label{sec:software}

\begin{figure*}[!t]
  \centering
  \includegraphics[width=5.25in, keepaspectratio]{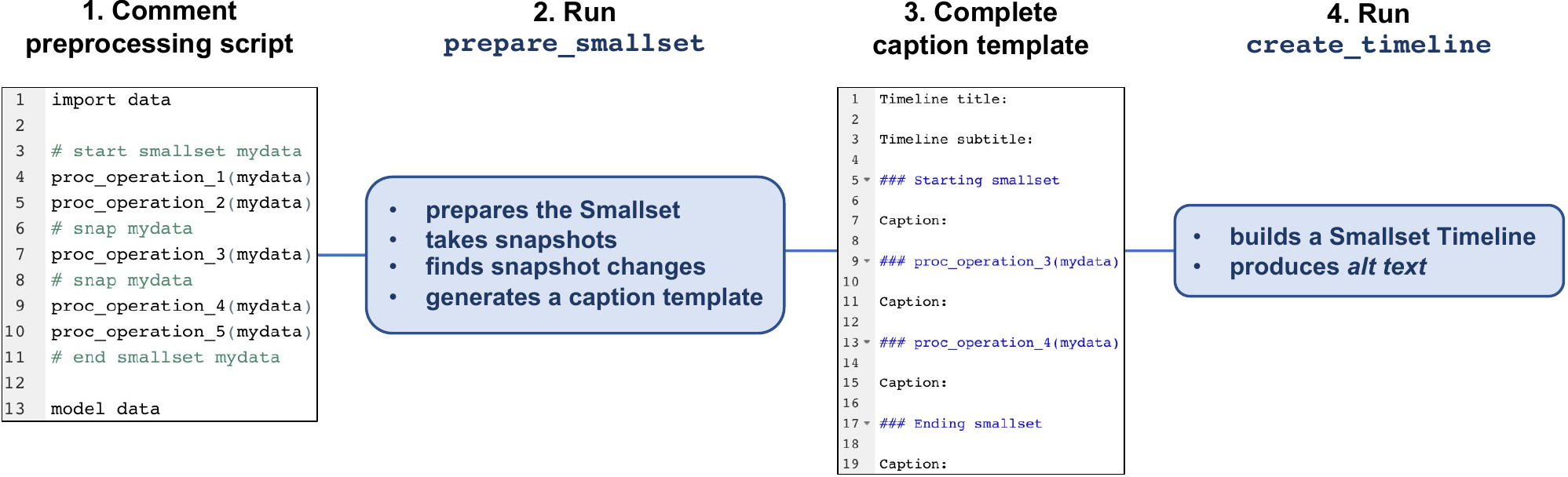}
  \caption{Steps for creating a \SsT using the \ssR software. Steps 1 and 3 requires \Timeline creator input of structured comments and captions, respectively. In Steps 2 and 4, \sstool takes data snapshots and builds the \Timeline. Example \sstool comments and the \texttt{R Markdown} caption for the synthetic data are in \Cref{app:code_for_syn}.}
  \label{fig:workflow}
\end{figure*}

All \SsTs presented in this work are made with the \sstool tool. 
\citet{Miceli2021} provide qualitative evidence that suggests producing documentation often feels like a ``burden'' to data practitioners. 
We develop a tool that aims to alleviate the burden and in turn encourage production of data preprocessing documentation. 
To integrate with existing or new preprocessing workflows, the software requires two inputs from \Timeline creators: adding structured comments to an R or Python preprocessing script and populating an R Markdown caption template generated by \sstool. 
\Cref{fig:workflow} contains a procedural overview of these inputs along with the \sstool processing tasks.

Producing a \SsT begins with a \Timeline creator adding a series of \sstool comments to an R or Python preprocessing script (\Cref{fig:workflow} Step 1).
Incorporating docstrings and comments to generate documentation is a common technique (e.g., \citep{Goodger2001, Wickham2020}). 
It is used here to assist in generating visual documentation of data preprocessing. 
The added comments provide instructions for \sstool, advising it where to take snapshots of the data or insert a resume marker (\Cref{ssec:op_features}). 
Each comment consists of one of four actions -- \textit{start smallset}, \textit{resume}, \textit{end smallset}, or \textit{snap} -- and the variable storing the data frame (e.g., \textit{\# snap mydata}). 
In Step 2, the software prepares the \Smallset, takes snapshots based on Step 1 input, analyses the snapshots for data changes, and generates a customised R Markdown caption template. 
Step 3 requires \Timeline creators to populate this template with captions for the snapshots. 
The caption input is used by \sstool as it assembles the \Timeline and produces the \alttext (\Cref{ssec:alt_text}) in Step 4. 
\Timeline creators can specify their preferences regarding \Smallset properties -- e.g., selection method (\Cref{sec:smallset_selection}) and size -- and \Timeline design -- e.g., colours, font, and enrichment features (\Cref{ssec:op_features}) -- in Steps 2 and 4, respectively.

We chose to do the initial implementation of \SsTs in R because it is a popular programming language for preprocessing datasets and offers strong graphics capabilities. 
We have enabled the software to also accept scripts in Python, another popular preprocessing choice. 
Future software development work can include increasing the capacity of \sstool to manage more complex preprocessing workflows, involving multiple scripts and the merging and joining of datasets.

\section{Case studies}
\label{sec:case_studies}

We present two case studies to illustrate the use of \SsTs. 
The first is on software defect detection data from the NASA Metrics Data Program (MDP). 
Despite being widely used to develop defect classification models, a lack of consistency in data preprocessing and documentation has jeopardised the utility of research outputs \citep{Gray2011, Gray2012, Shepperd2013, Petric2016}. 
The second case study examines benchmark datasets containing American Community Survey (ACS) data. 
We quantify differences in fairness metrics due to different preprocessing decisions, whereas recent work \citep{Ding2021} has focused on differences in fairness metrics across fairness interventions and income thresholds.

\subsection{NASA MDP data}
\label{ssec:nasamdp}

\begin{figure*}[!t]
  \centering
  \includegraphics[width=\linewidth, keepaspectratio]{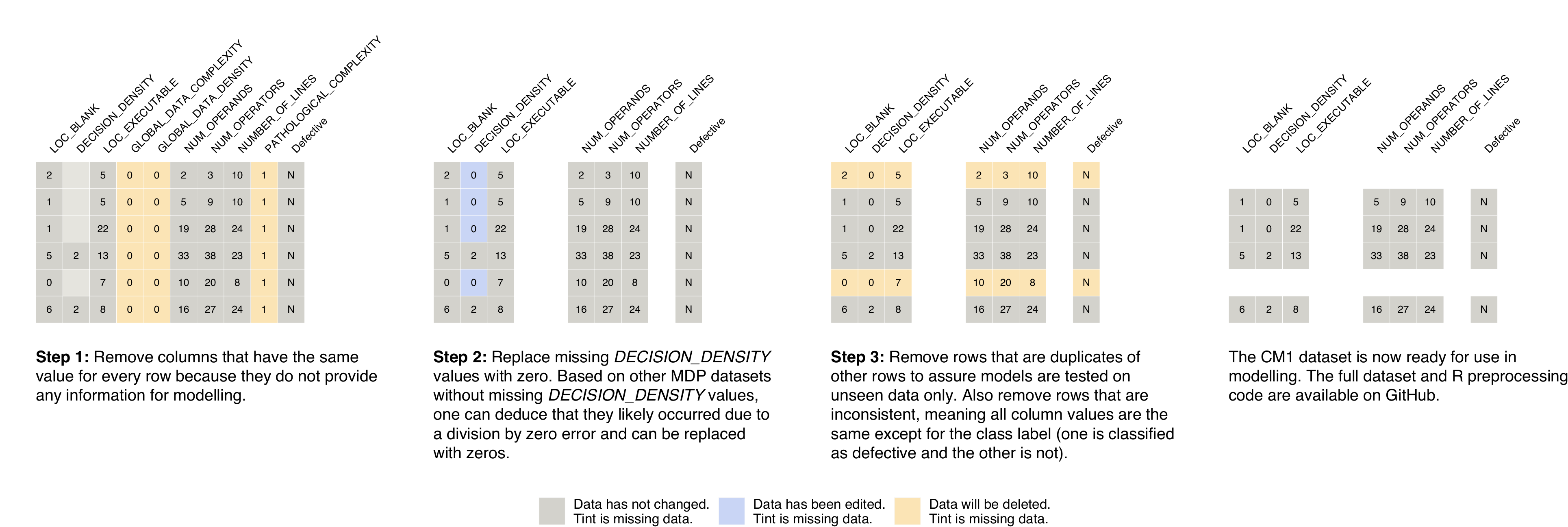}
  \caption{\SsT for MDP CM1 dataset preprocessed according to \citep{Gray2011}. See \Cref{ssec:mdp_understanding} for discussion. \Smallset selected using Problem 1 algorithm (see \Cref{sec:smallset_selection}).}
  \label{fig:gray_general}
\end{figure*}

In the early 2000s, the MDP released 13 datasets for software defect detection research. 
Like many real-world datasets, the data require preprocessing. 
There are missing, erroneous, extraneous, and duplicate data to address. 
We chose these data as a case study because of existing literature \citep{Gray2011, Gray2012, Shepperd2013, Petric2016} focused on assessing MDP preprocessing practices. 
For example, \citet{Gray2012} note the issue of duplicate data occurring in the testing and training set (i.e., the model is not tested on unseen data). Their concern is that ``the impression given from the literature is that many defect prediction researchers using this data have not been aware of this issue'' \citep[p.~557]{Gray2012}. 
\citet{Shepperd2013} highlight that studies deal with the data issues differently and are ``not in the habit of providing complete information regarding preprocessing of data'' [p.~1213]. 
The literature presents a clear example of insufficient documentation for preprocessing decisions, i.e., an example of the problem that \SsTs are designed to address. 
In the rest of this section, we use the MDP CM1 dataset \citep{Tantithamthavorn2016}. It has 505 rows and 41 columns. 
For the \SsTs in \Cref{ssec:mdp_understanding} and \Cref{ssec:mdp_rep}, we choose to display 10 and 15 columns, respectively.

\subsubsection{\SsTs for comprehension}
\label{ssec:mdp_understanding}

Figure~\ref{fig:gray_general} contains a \SsT for the preprocessing strategy recommended by \citet{Gray2011}.
The \Timeline uses a \Smallset with six rows and consists of four snapshots. 
It discloses how the common MDP data issues have been dealt with. For instance, snapshot 2 discusses missing data. 
In some work, they are simply dropped~\citep{Shepperd2013}. 
However, this \Timeline creator argues that, based on study of other MDP datasets, the missing values can be attributed to a division by zero error and retained by imputing zeros \citep{Gray2011}.
Snapshot 3 addresses the issue of duplicate data, noting that it is removed and why this is necessary.
\Cref{fig:gray_general} uses about the same amount of space as the other figures in this work (which are not \SsTs, e.g., \Cref{fig:definition} or \Cref{fig:workflow}) and remains legible.

\subsubsection{\SsTs for replication}
\label{ssec:mdp_rep}

\begin{figure*}[!t]
  \centering
  \includegraphics[width=5.75in, keepaspectratio]{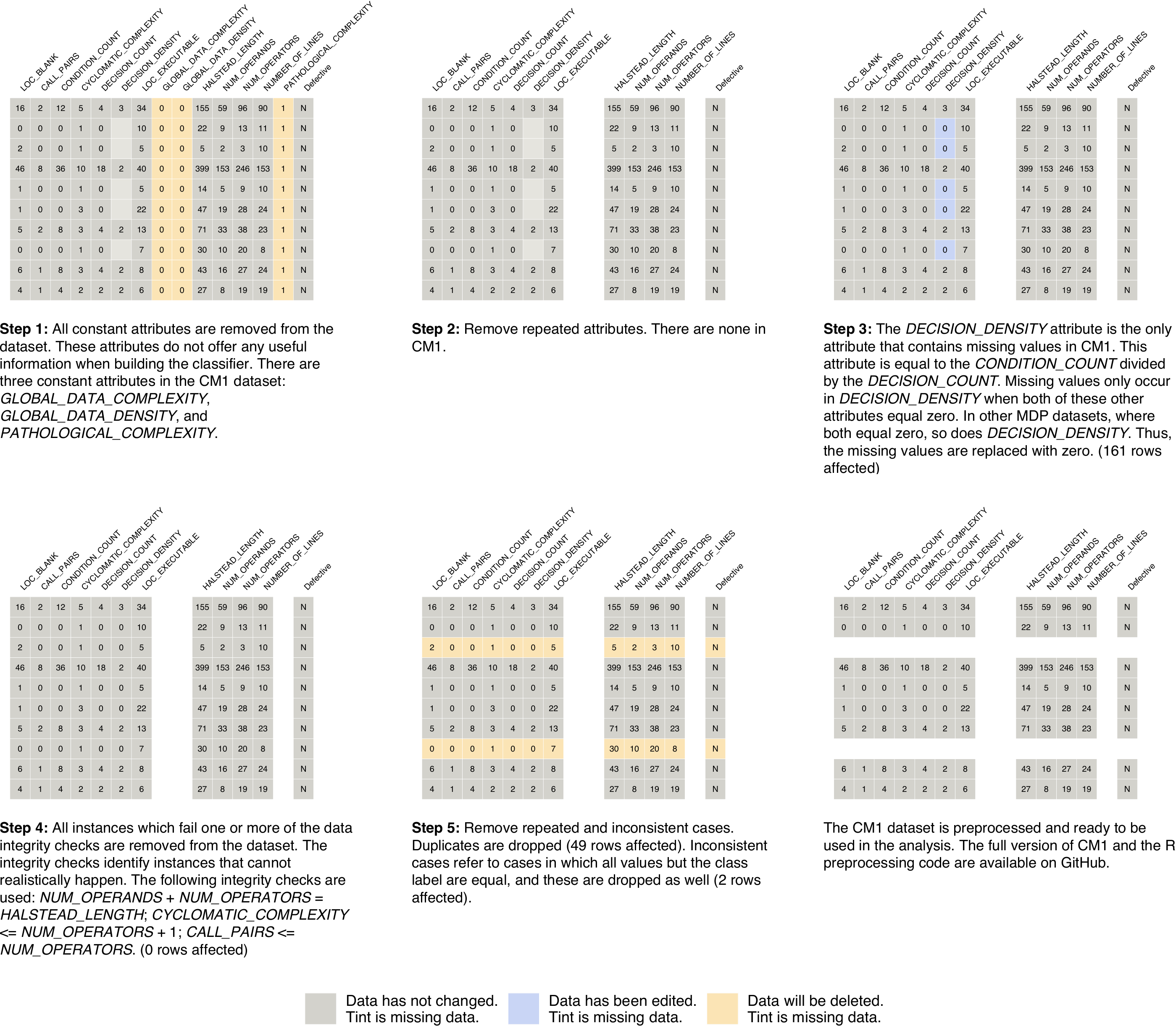}
  \caption{\SsT for MDP CM1 dataset, for replication. See \Cref{ssec:mdp_rep} for discussion. \Smallset selected using Problem 1 algorithm (see \Cref{sec:smallset_selection}).}
  \label{fig:gray_replication}
\end{figure*}

In this section, we restructure the \SsT presented in \Cref{fig:gray_general} to better support replication efforts (\Cref{fig:gray_replication}, \Cref{table:functions}). 
\citet{Shepperd2013} suggest that practitioners using MDP data ``report any preprocessing in sufficient detail to enable meaningful replication'' [p.~1208]. 
For replication, the \Timeline needs to be comprehensive and specific. 
We increase the \Smallset size to include additional data examples and take more snapshots to separate the process into its component parts. 
As a result, the \Timeline is larger and may be located in an appendix or with the preprocessing code. 
The captions contain specific information, including the total number of rows an operation affects and the rules for checking data integrity.

Data integrity checks are an important MDP preprocessing step that remove implausible values. 
As noted in caption 4 of \Cref{fig:gray_replication}, the checks do not actually affect any rows in the CM1 dataset. 
The step was left out of \Cref{fig:gray_general} for brevity, but it is included here for clarity. 
If replicating the preprocessing strategy on another dataset, it would be necessary for accuracy and consistency to conduct the data checks. 
It is worth noting that the preprocessing strategy proposed in \citet{Shepperd2013} suggests running 18 different integrity checks, while \citet{Petric2016} suggest 20 different integrity checks. 
The additional checks do result in the loss of observations in CM1. 
In other words, indicating that  ``the data checks were run'' is not enough information. 
Replication will require greater specification.

\subsection{The \texttt{folktables} data}
\label{ssec:folktables}

The UCI Adult dataset \citep{kohavi1996} consists of 1994 census income data, and the associated estimation task is to predict if an individual earns more than 50,000 dollars per year. 
It has been used in hundreds of research papers related to machine learning fairness.
A recent paper by \citet{Ding2021} challenges the machine learning community’s continued reliance on the dataset, given its age and defects.
For example, the 50,000 dollar threshold leads to imbalance by race and gender in the dataset as it represents the ``88th quantile in the Black population, and the 89th quantile among women'' \citep[p.~2]{Ding2021}.
In turn, they develop a tool, \texttt{folktables}, to generate recent benchmark datasets from the American Community Survey (ACS) and define new prediction tasks.
It allows adjustable income thresholds and data filtering criteria.
We explore effects of these preprocessing decisions on 2015 ACS income data from California (CA), Connecticut (CT), and Utah (UT), retrieved with \texttt{folktables}.

\begin{figure*}[!t]
  \centering
  \includegraphics[width=5.5in, keepaspectratio]{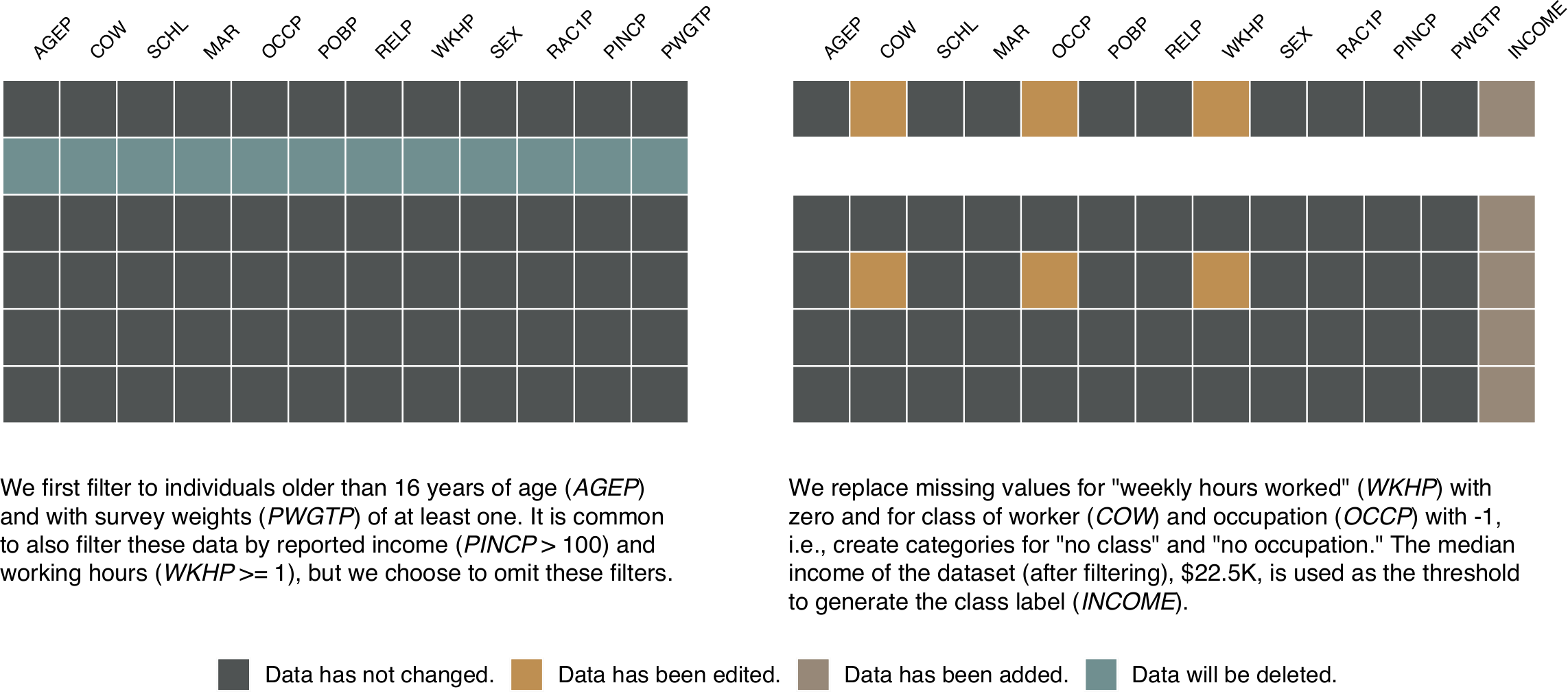}
  \caption{\SsT of ACS California data preprocessed with the \textit{validity-median} setting.
  \Smallset selected with random sampling.
  See \Cref{ssec:folktables} for discussion and \Cref{app:python} for the Python preprocessing script behind this \Timeline.}
  \label{fig:census}
\end{figure*}

\begin{figure*}[!t]
  \centering
  \includegraphics[width=5.6in, keepaspectratio]{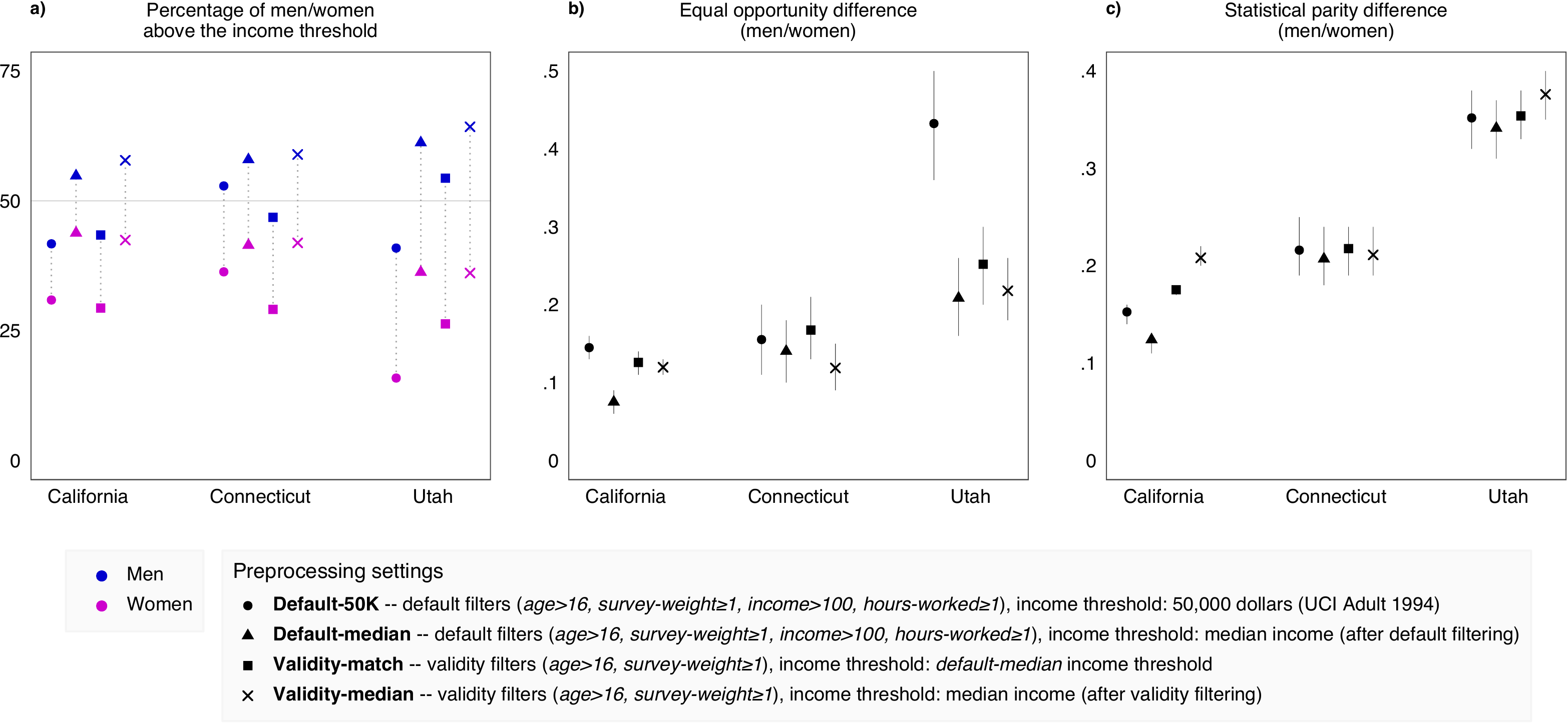}
  \caption{The effect of four different preprocessing settings on data and prediction. 
  (a) Dataset imbalance by sex. 
  (b) and (c) Group fairness measures in predictions, error bars refer to 95\% Newcombe intervals. 
  See \Cref{ssec:folktables} for discussions.}
  \label{fig:fairness}
\end{figure*}

We explore four different preprocessing scenarios, starting with the default setting used by \citet{Ding2021} -- referred to here as \textit{default-50K}.
In this setting, an income threshold of \$50K is applied to generate positive and negative labels, after filtering the dataset to retain an individual's record when they 1) are older than 16 years of age, 2) have a survey weight of at least one, 3) earn more than 100 dollars, and 4) report at least one hour of usual weekly work. 
The next setting, called \textit{default-median}, uses the same set of default data filters but sets the income threshold to the sample median after filtering (\$36K, \$45K, and \$31.1K for CA, CT, and UT, respectively) to generate more balanced prediction tasks.
The remaining settings aim to be inclusive of all the target population, on the grounds that individuals who did not work and/or reported income losses are still valid instances for prediction, by dropping the last two filters.
We refer to this filtering approach as ``validity.''
The setting \textit{validity-match} uses the same threshold as \textit{default-median}, such that the thresholds ``match'' (e.g., \$36K for CA) despite different filters.
Lastly, the \textit{validity-median} setting uses its own sample median after validity filtering (\$22.5K, \$30.2K, and \$23.5K for CA, CT, and UT, respectively).
\Cref{fig:census} is an example \SsT depicting the \textit{validity-median} steps applied to the California dataset. Additional dataset statistics are available in \Cref{app:data_info}.

Figure~\ref{fig:fairness} presents results across preprocessing settings on class imbalance in gender\footnote{In the original dataset, the attribute corresponds with a male/female encoding and does not include nonbinary gender options.} and fairness levels in classification results. 
\Cref{fig:fairness}(a) compares the percentage of men and women above the income threshold.
It shows that certain income thresholds achieve greater balance than others, e.g., for California, \textit{default-median} is more balanced among groups (aiming for equal splits overall) than \textit{default-50K}.
However, using the same threshold alone will not guarantee balance or consistency across studies that use different preprocessing filters.
For example, comparing \textit{default-median} and \textit{validity-match} for California, which have matching thresholds (\$36K) but different filters, we see a substantial change in the percentage of women above the income threshold (43.8\% and 29.4\% for \textit{default-median} and \textit{validity-match}, respectively).
Thus, it is necessary to communicate both filtering decisions and the threshold selection.

With \texttt{folktables}, we define prediction tasks that correspond with \textit{default-50K}, \textit{default-median}, \textit{validity-match}, and \textit{validity-median}.
All tasks predict if an individual's income is over the threshold.
For each task and state, we train and test a logistic regression model (with \texttt{scikit-learn} \citep{scikit-learn} default settings) on 80\% and 20\% of the dataset, respectively.
For men and women in the dataset, we compute differences of equality of opportunity (EO) \citep{Hardt2016} and statistical parity (SP) \citep{Dwork2012} from the test set predictions and 95\% Newcombe intervals \citep{Newcombe1998} for the differences.
\Cref{fig:fairness}(b) shows a significant difference in the EO between \textit{default-50K} and the other three settings for Utah, but variations among different settings in Connecticut are much smaller.
\Cref{fig:fairness}(c) shows that, across the four settings, the SP values are significantly different for California but not for Connecticut or Utah.

\section{Conclusion}
We present the \SsT, a visualisation of data preprocessing decisions.
It is designed to support reflection by \Timeline creators and replication, comprehension, and evaluation by \Timeline readers.
Its static, compact nature makes it a practical figure to include in  research outputs.
We develop the \sstool tool, an R software package for producing \SsTs from R and Python scripts.
\Timeline creators only need to add a few structured comments to the preprocessing script and supply captions.
We also present case studies on software defect data and income survey benchmark data, highlighting the importance of communicating decisions that affect the dataset and prediction outcomes.
We include several \SsTs to illustrate use of the visualisation and software tool.

Future work involves new features, visual design, and software development.
Examples include: to incorporate data statistics in diagrams alongside the \Smallset snapshots, to support {\em comparison} between different preprocessing decisions, and to design succinct visualisations for complex workflows such as dataset joins and a richer set of data formats.
We are also interested in techniques for assuring data privacy in a \Smallset, visual representation for specific preprocessing tasks, and new applications.
Lastly, it will be great to potentially incorporate \SsTs within other data science provenance tools.
For instance, a \SsT could be included as part of Question 33 in datasheets~\citep{gebru2021datasheets}, in the {\em Evaluation data} section of model cards~\citep{Mitchell2019}, and the dataset composition section of Dataset Nutrition Labels~\citep{Holland2018}.

\begin{acks}
This research is supported in part by the Australian Research Council Project DP180101985. 
The first author is supported in part by a CSIRO's Data61 Top-Up scholarship. 
We thank the ANU Humanising Machine Intelligence team, C\'ecile Paris, and the reviewers for their comments and thoughtful suggestions.
\end{acks}

\bibliographystyle{ACM-Reference-Format}
\bibliography{main}

\clearpage
\onecolumn
\appendix

\section{Synthetic Dataset}
\label{app:synthetic}

For illustrative purposes, we generate a synthetic dataset and preprocessing scenario. 
This section describes its basic profile (\Cref{app:data}), the three preprocessing steps it undergoes (\Cref{app:preproc}), a \SsT for these steps (\Cref{app:timeline}), another \Timeline with a resume marker (\Cref{app:resume}), \alttext for the \SsT in \Cref{app:timeline} (\Cref{app:alt_output}), the data matrices for the \Smallset selection algorithms (\Cref{app:score_matrices}), and the R preprocessing script with structured comments and populated caption template passed to the \sstool software (\Cref{app:code_for_syn}).

\subsection{Dataset}
\label{app:data}

\begin{figure*}[h]
\centering
\includegraphics[width=2.5in, keepaspectratio]{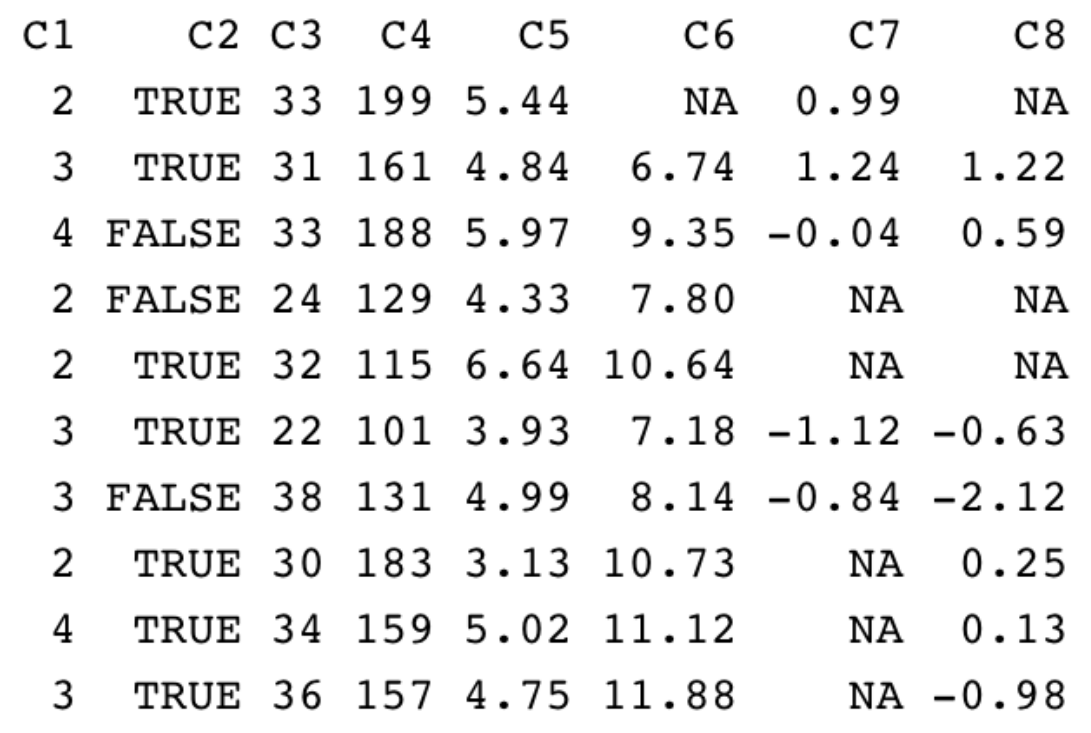}
\caption{First ten rows of the synthetic dataset.}
\label{fig:synth_rows}
\end{figure*}

The dataset is synthesised with the \texttt{charlatan} software package \citep{Chamberlain2020} in R. 
The initial dataset consists of 100 rows and 8 features. The features are described in \Cref{table:synthetic}, and the first ten rows of the dataset are printed in \Cref{fig:synth_rows}.
\begin{table*}[h]
\small
\begin{tabular}{ccc}
    \toprule Name & Type & Missing Values \\ \midrule
    C1 & Categorical &  No \\
    C2 & Binary &  No \\
    C3 & Discrete & No \\
    C4 & Discrete & No \\
    C5 & Continuous &  No \\
    C6 & Continuous &  Yes (14\%) \\
    C7 & Continuous &  Yes (44\%) \\
    C8 & Continuous &  Yes (19\%) \\
 \bottomrule
\end{tabular}
\caption{Features descriptions for the synthetic dataset.}
\label{table:synthetic}
\end{table*}

\subsection{Preprocessing}
\label{app:preproc}

The data preprocessing for the synthetic dataset consists of three main steps.
\begin{enumerate}
    \item Filter rows
    \begin{itemize}
        \item Remove rows where C2 is FALSE
    \end{itemize}
    \item Deal with missing data
    \begin{itemize}
        \item Replace missing values in C6 and C8 with mean values by C1 category
        \item Drop C7
    \end{itemize}
    \item Generate a new feature
    \begin{itemize}
        \item Create C9 by summing C3 and C4
    \end{itemize}
\end{enumerate}

\clearpage
\subsection{\SsT}
\label{app:timeline}

\begin{figure*}[h]
\centering
\includegraphics[width=\linewidth, keepaspectratio]{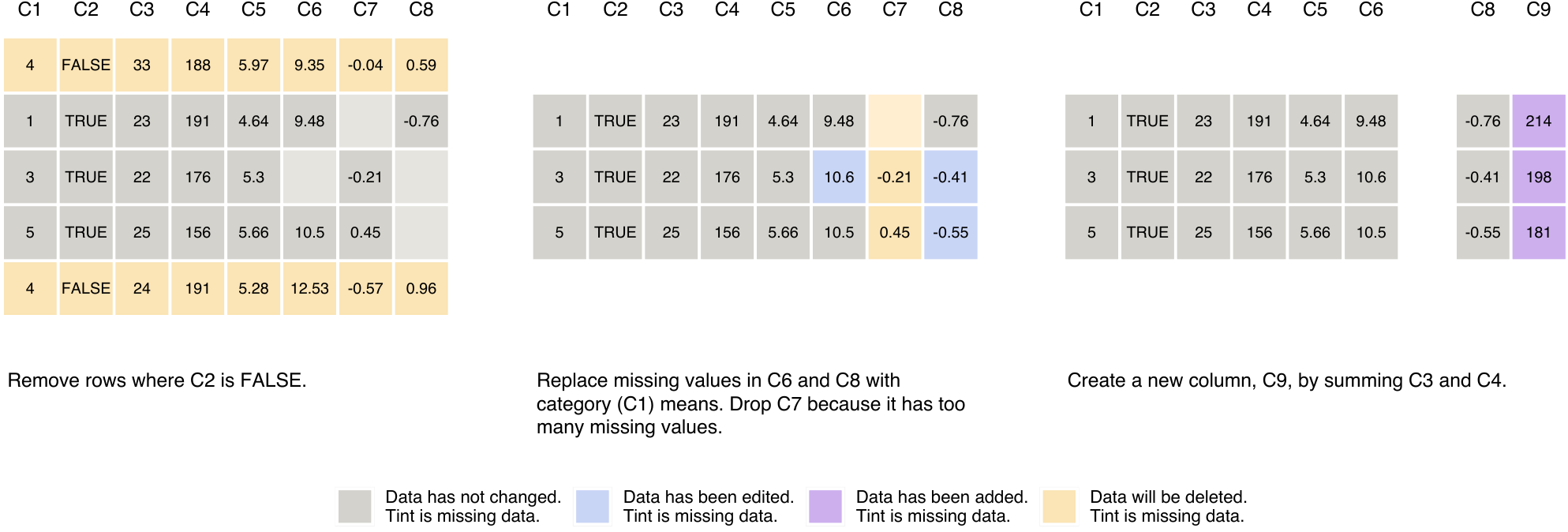}
\caption{\SsT for the synthetic dataset and the preprocessing scenario. \Smallset selected using Problem 2 algorithm.}
\label{fig:synth_timeline}
\end{figure*}

Figure~\ref{fig:synth_timeline} is a \SsT for the synthetic data example.

\subsection{Resume markers}
\label{app:resume}

\begin{figure*}[h]
\centering
\includegraphics[width=\linewidth, keepaspectratio]{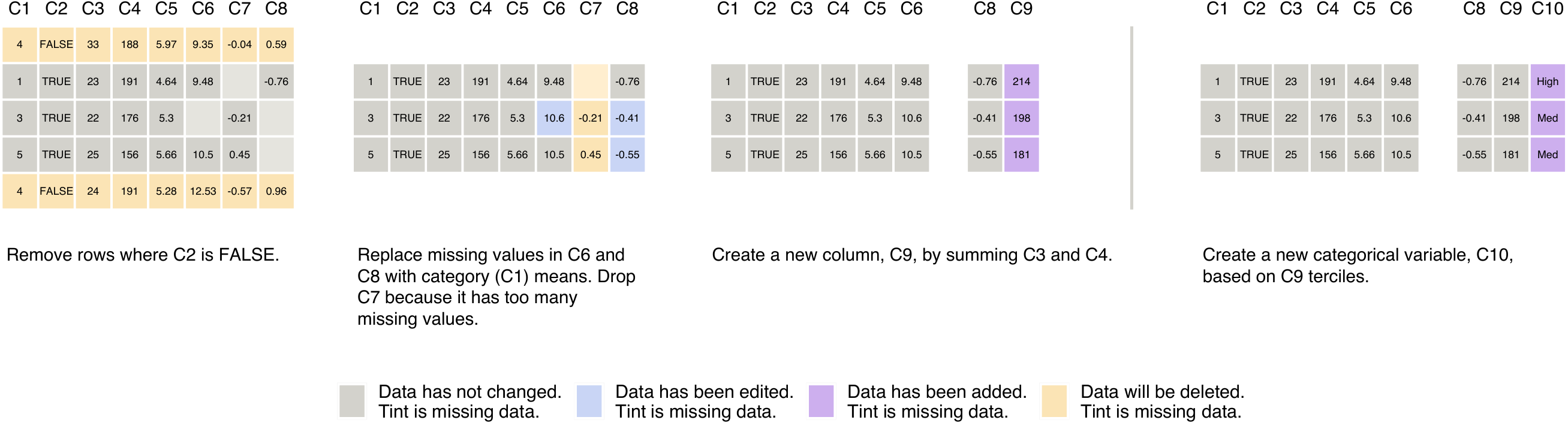}
\caption{\SsT with a resume marker.}
\label{fig:resume}
\end{figure*}

Figure~\ref{fig:resume} includes an additional step (generation of another feature) in the \Timeline to illustrate use of the \textbf{resume markers} enrichment feature.

\clearpage
\subsection{\Alttext}
\label{app:alt_output}

\begin{figure*}[h]
\centering
\includegraphics[height = 3in, keepaspectratio]{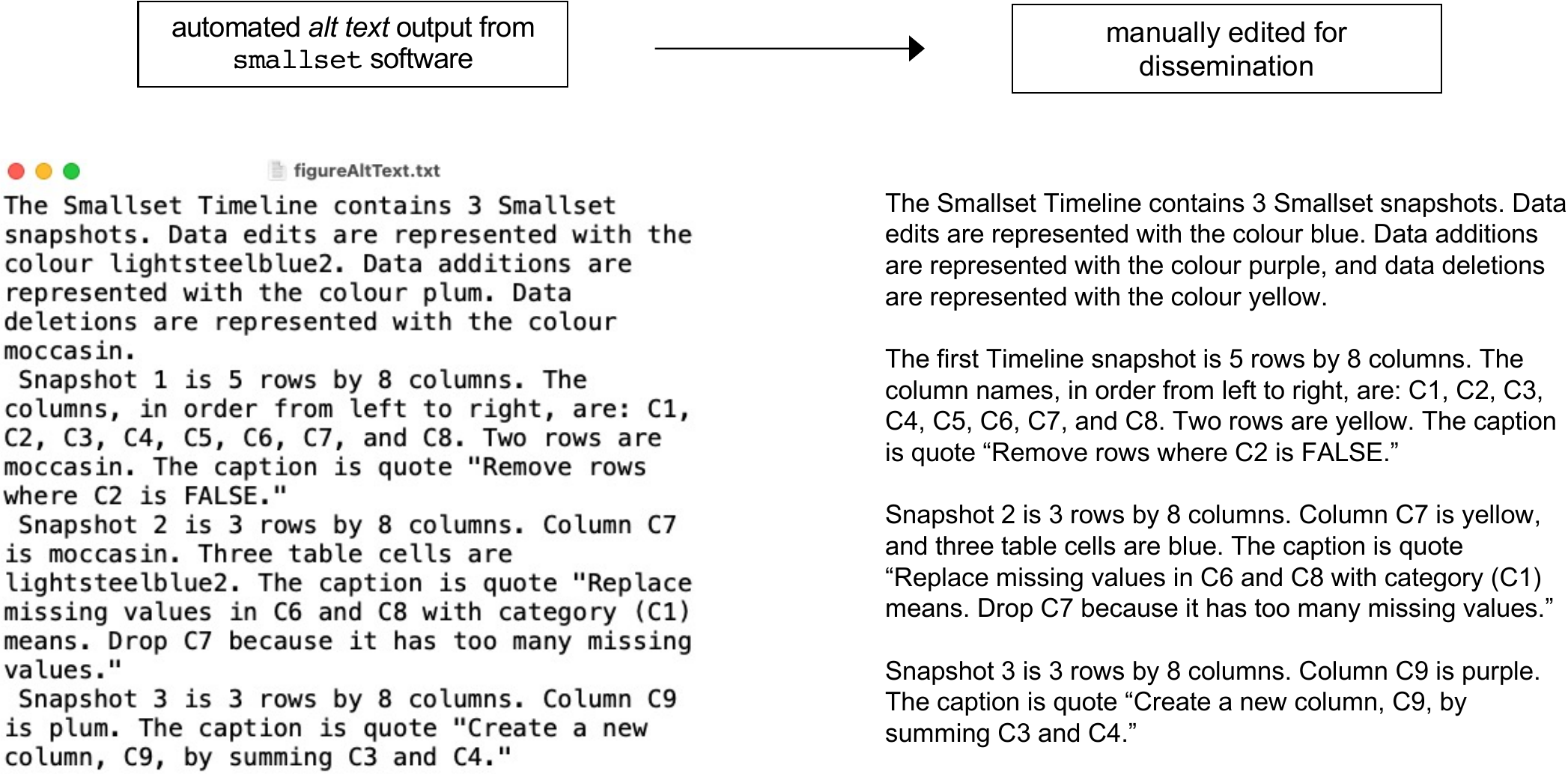}
\caption{Example of automated \alttext generated by the \sstool software (left) and a manually edited version of it prepared for dissemination (right). The \alttext is a description of the \SsT in \Cref{fig:synth_timeline}.}
\label{fig:output}
\end{figure*}

\subsection{Data representations for \Smallset selection}
\label{app:score_matrices}

\begin{figure*}[h]
\[
\begin{blockarray}{cccc}
  & Step1 & Step2 & Step3 \\
 \begin{block}{c[ccc]}
 1 & 0 & 1 & 1 \\
 2 & 0 & 1 & 1 \\
 3 & 1 & 0 & 0 \\
 4 & 1 & 0 & 0 \\
 5 & 0 & 1 & 1 \\
 & \vdots & \vdots & \vdots\\
 98 & 0 & 1 & 1 \\
 99 & 1 & 0 & 0 \\
 100 & 0 & 1 & 1 \\
 \end{block}
 \end{blockarray}\vspace*{-1.25\baselineskip}
\qquad
 \begin{blockarray}{cccccccccc}
  & C1 & C2 & C3 & C4 & C5 & C6 & C7 & C8 & C9 \\
 \begin{block}{c[ccccccccc]}
 1 & \gr & \gr & \gr & \gr & \gr & \bl & \ye & \bl & \pu \\
 2 & \gr & \gr & \gr & \gr & \gr & \gr & \ye & \gr & \pu \\
 3 & \ye & \ye & \ye & \ye & \ye & \ye & \ye & \ye & \ye \\
 4 & \ye & \ye & \ye & \ye & \ye & \ye & \ye & \ye & \ye \\
 5 & \gr & \gr & \gr & \gr & \gr & \gr & \ye & \bl & \pu \\
  & \vdots & \vdots & \vdots & \vdots & \vdots & \vdots & \vdots & \vdots & \vdots\\
 98 & \gr & \gr & \gr & \gr & \gr & \gr & \ye & \gr & \pu \\
 99 & \ye & \ye & \ye & \ye & \ye & \ye & \ye & \ye & \ye \\
 100 & \gr & \gr & \gr & \gr & \gr & \gr & \ye & \gr & \pu \\
 \end{block}
 \end{blockarray}
\]
\caption{\textit{Coverage indicator matrix} (left) and \textit{visual appearance matrix} (right) for the synthetic data example. The letters in the \textit{visual appearance matrix} represent the change last affecting a cell (\textit{\textbf{U}}: unchanged, \textit{\textbf{E}}: edit, \textit{\textbf{A}}: addition, \textit{\textbf{D}}: deletion).}
\label{fig:matrices}
\end{figure*}

\clearpage
\subsection{{\sstool} software input for synthetic data}
\label{app:code_for_syn}

\begin{figure*}[h]
\centering
\includegraphics[width=5in, keepaspectratio]{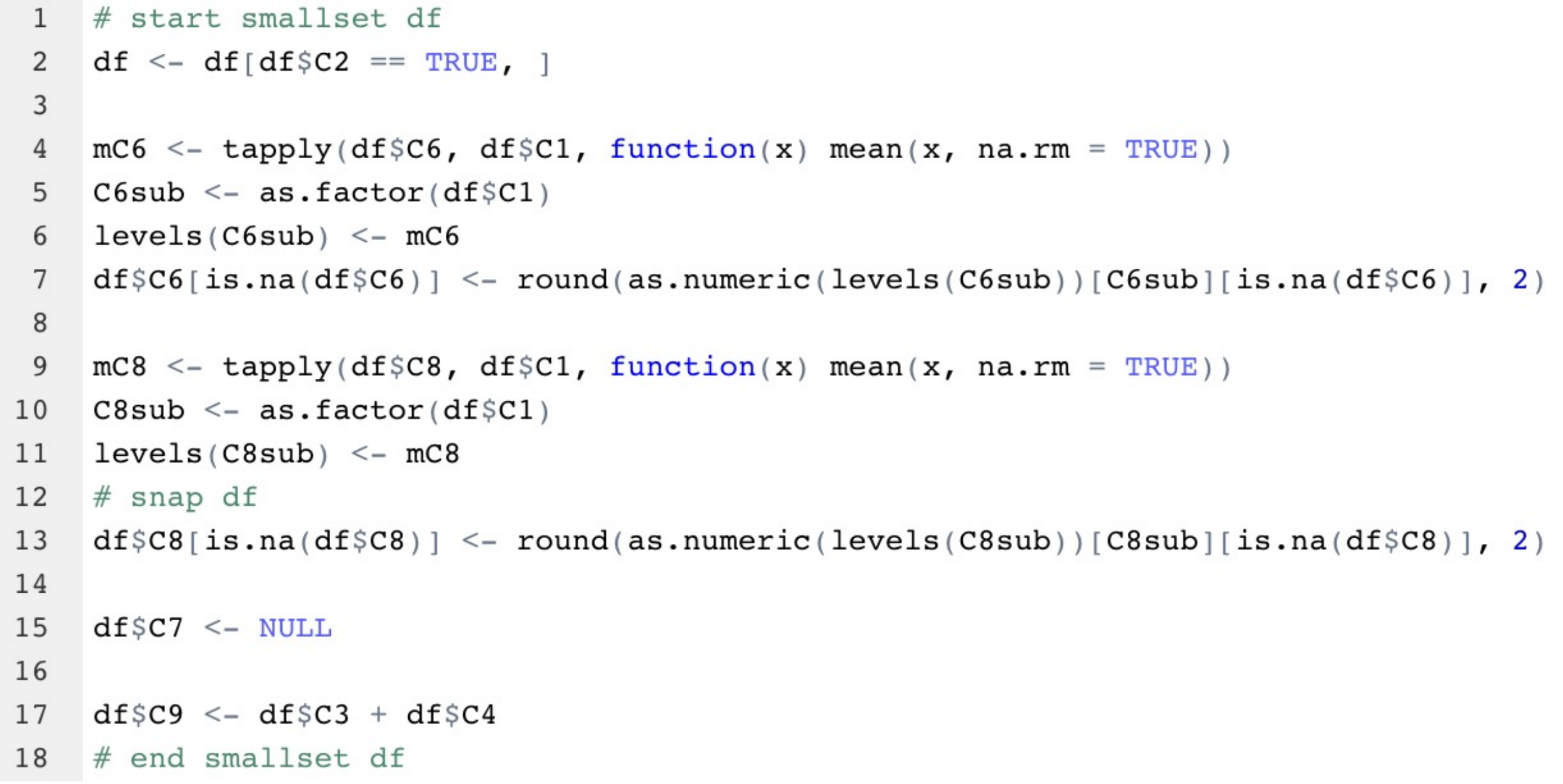}
\caption{Structured \sstool comments in the R preprocessing script for the synthetic dataset and \SsT in \Cref{fig:synth_timeline}.}
\label{fig:comments}
\end{figure*}

\begin{figure*}[h]
\centering
\includegraphics[width=5in, keepaspectratio]{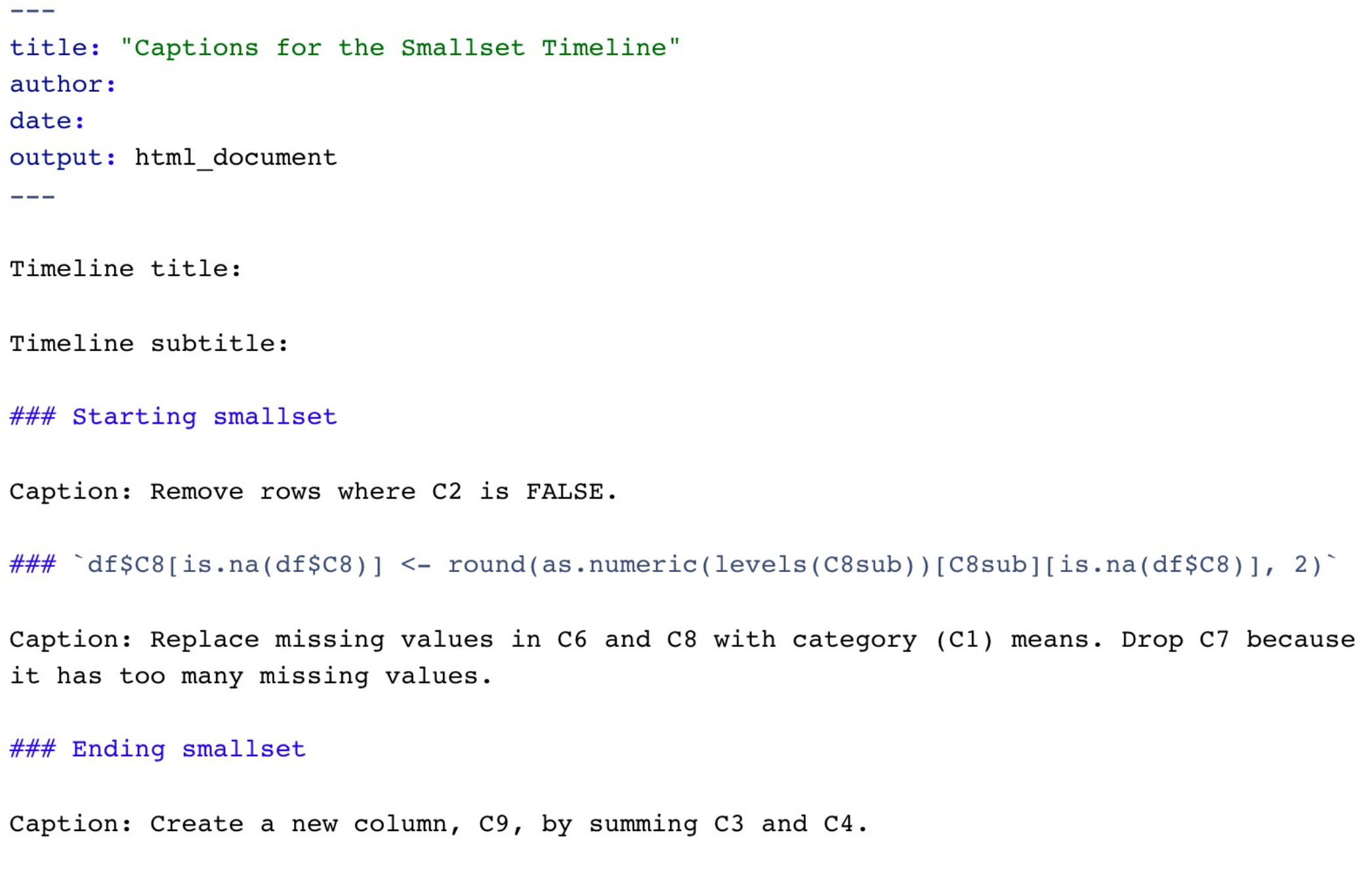}
\caption{Completed R Markdown caption template for the synthetic dataset and \SsT in \Cref{fig:synth_timeline}.}
\label{fig:captions}
\end{figure*}

\clearpage
\section{Alt text template}
\label{app:alt_template}

\begin{figure*}[h]
\centering
\includegraphics[height = 4in, keepaspectratio]{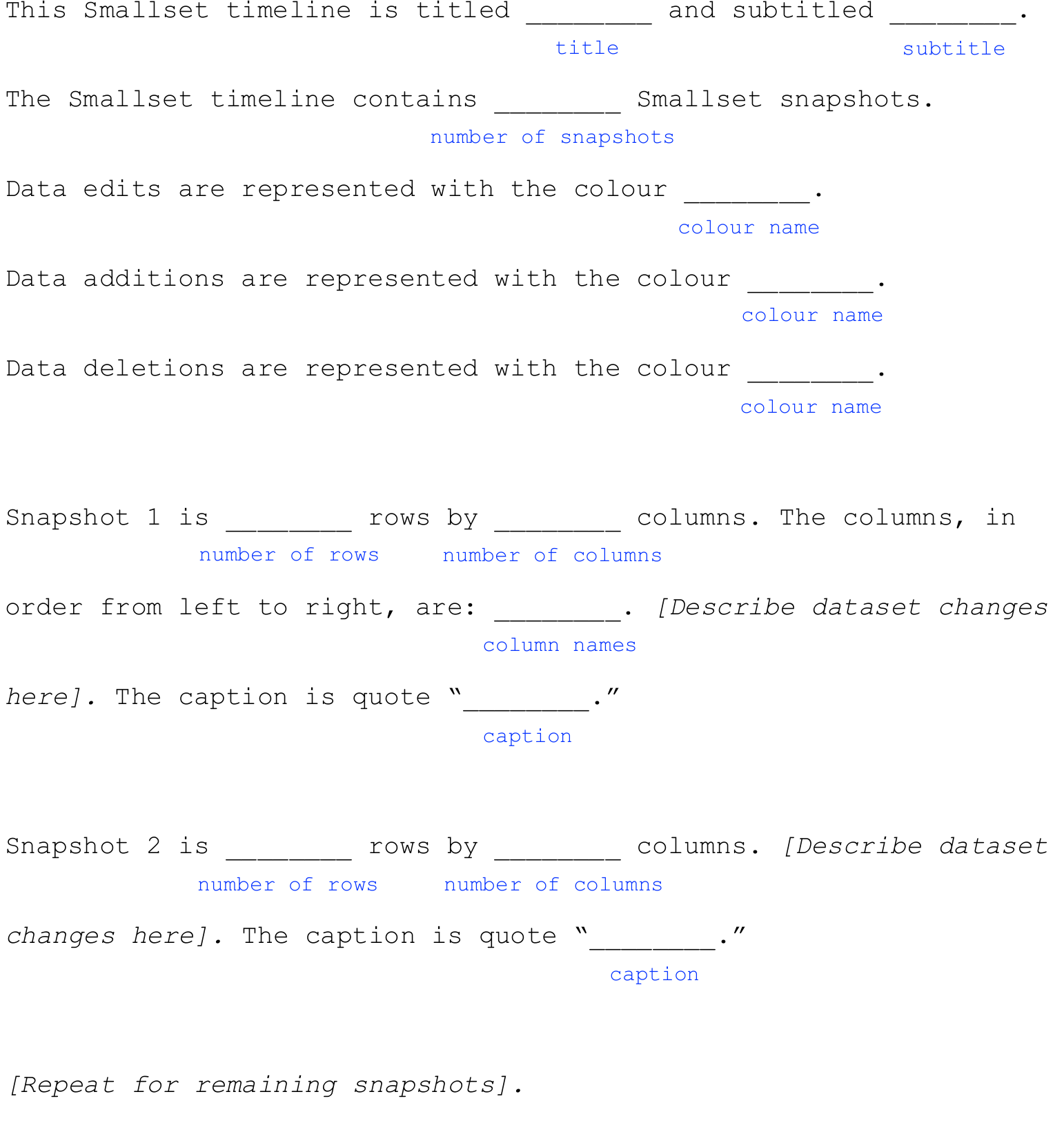}
\caption{\Alttext template for \SsTs.}
\label{fig:template}
\end{figure*}

\clearpage
\section{folktables data}
\label{app:folktables}

\subsection{Preprocessing script}
\label{app:python}

\begin{figure*}[h]
\centering
\includegraphics[height = 1in, keepaspectratio]{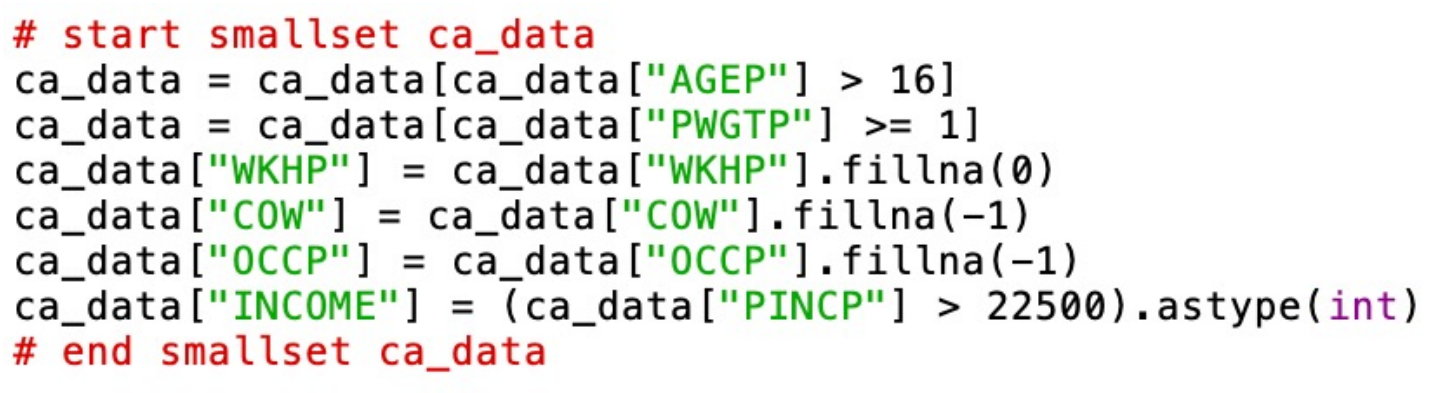}
\caption{Python preprocessing script for the \SsT in \Cref{fig:census}. No intermediary snapshot points are specified with a ``\texttt{\# snap ca\_data}'' comment, resulting in a two-snapshot \Timeline. This script does not mirror the exact \texttt{folktables} preprocessing workflow but does execute the same preprocessing operations and demonstrates the capacity of \sstool to accept Python scripts. Future work includes increasing the capacity of \sstool to handle different workflow styles, such as the one used in \texttt{folktables}, which includes calling/called functions.}
\label{fig:python_script}
\end{figure*}

\subsection{Dataset information}
\label{app:data_info}

\begin{table*}[h]
\small
\caption{Sample sizes and male/female counts for state datasets, before and after data filtering.}
\begin{tabular}{cccccccccc}
    \toprule  & 
    \multicolumn{3}{c}{Before filtering} & 
    \multicolumn{3}{c}{After default filtering} & 
    \multicolumn{3}{c}{After validity filtering}\\ \cmidrule(lr){2-4}\cmidrule(lr){5-7}\cmidrule(lr){8-10}
      & Total (\textit{n}) & Males & Females & Total (\textit{n}) & Males & Females & Total (\textit{n}) & Males & Females\\
      \midrule
      California & 374,943 & 184,637 & 190,306 & 187,475 & 99,518 & 87,957 & 299,619 & 146,131 & 153,488 \\
      Connecticut & 35,787 & 17,270 & 18,517 & 19,398 & 9,926 & 9,472 & 29,232 & 13,868 & 15,364 \\
      Utah & 29,290 & 14,614 & 14,676 & 14,868 & 8,174 & 6,694 & 21,235 & 10,439 & 10,796 \\
     \bottomrule
\end{tabular}
\end{table*}

\end{document}